# BlueWalker 3 Satellite Brightness Characterized and Modeled

Anthony Mallama, Richard E. Cole, Scott Tilley, Cees Bassa and Scott Harrington

Contact: anthony.mallama@gmail.com

Abstract

The BlueWalker 3 (BW3) satellite was folded into a compact object when launched on 2022 September 11. The spacecraft's apparent visual magnitude initially ranged from about 4 to 8. Observations on November 11 revealed that the brightness increased by 4 magnitudes which indicated that the spacecraft had deployed into a large flat-panel shape. The satellite then faded by several magnitudes in December before returning to its full luminosity; this was followed by additional faint periods in 2023 February and March. We discuss the probable cause of the dimming phenomena and identify a geometrical circumstance where the satellite is abnormally bright. The luminosity of BW3 can be represented with a brightness model which is based on the satellite shape and orientation as well as a reflection function having Lambertian and pseudo-specular components. Apparent magnitudes are most frequently between 2.0 and 3.0. When BW3 is near zenith the magnitude is about 1.4.

## 1. Introduction

BlueWalker 3 (BW3) unfolded into an extremely large and bright satellite on-orbit in 2022 November. A press release issued by the International Astronomical Union describes the adverse affect that BW3 and the expected constellation of similar spacecraft will have on astronomical research and on the aesthetic appearance of the night sky.

This paper chronicles the visual brightness of BW3 from its launch on 2022 September 11 through 2023 April 19. Section 2 provides an overview of magnitude changes during several periods of time: before the satellite unfolded, after it unfolded and attained its regular brightness and three intervals of fading that occurred in 2022 December and 2023 February and March. (The term *regular* brightness is used for magnitudes obtained after unfolding and outside of the faint periods.) Section 3 characterizes the regular brightness in terms of its most frequently observed apparent magnitude, the brightness observed when the satellite is near zenith and the illumination phase curve. Section 4 describes a brightness model for BW3, and presents findings which apply to the satellite's regular luminosity and a few unusually bright pseudo-specular magnitudes. Section 5 gives the conclusions from this study and suggests how the adverse impact of BW3 on astronomy can be ameliorated.

This work is part of a growing area of research into the characteristics of artificial satellite and the problems they create for astronomy. Photometry by Halferty et al. (2022) and Mroz et al. (2022) documents their impact on scientific observations, while Tyson et al. (2020) analyze the potential interference with critical observations being planned for Rubin Observatory. A wide range of research topics that address satellite constellations can be found in Walker and Benvenuti (2022).

## 2. Observations and light curves

Magnitudes of BW3 were measured by comparing the satellite to nearby reference stars. Proximity between the satellite and those stars accounts for variations in sky transparency and sky brightness. Most luminosity measurements were obtained with the unaided eye or through binoculars using methods similar to those described by Mallama (2022). Other measurements were derived from CMOS and video imagery. The observers and their geodetic coordinates are listed in Appendix A and the observations are tabulated in Appendix B.

The light curve for BW3 with apparent magnitudes adjusted to a uniform distance is shown in Figure 1. Distinct changes in brightness indicate deployment in November, the fading periods in 2022 December and 2023 February and March, and recoveries to regular luminosity. The brightening associated with deployment was documented by Mallama et al. (2022) as was the first part of the December faint period by Mallama et al. (2023).

The three faint periods occurred when the satellite's orbital plane became nearly perpendicular to the solar direction. As a result, the normally zenith facing side of the flat-panel shaped spacecraft, which supports the solar array, would have been oriented edgewise to the Sun. So, it would receive very little insolation.

Mallama et al. (2023) hypothesized that satellite operators tilted the zenith side toward the Sun in order to supply sufficient power to run the spacecraft. Therefore, the nadir side, which is seen by observers on the ground, was mostly dark.

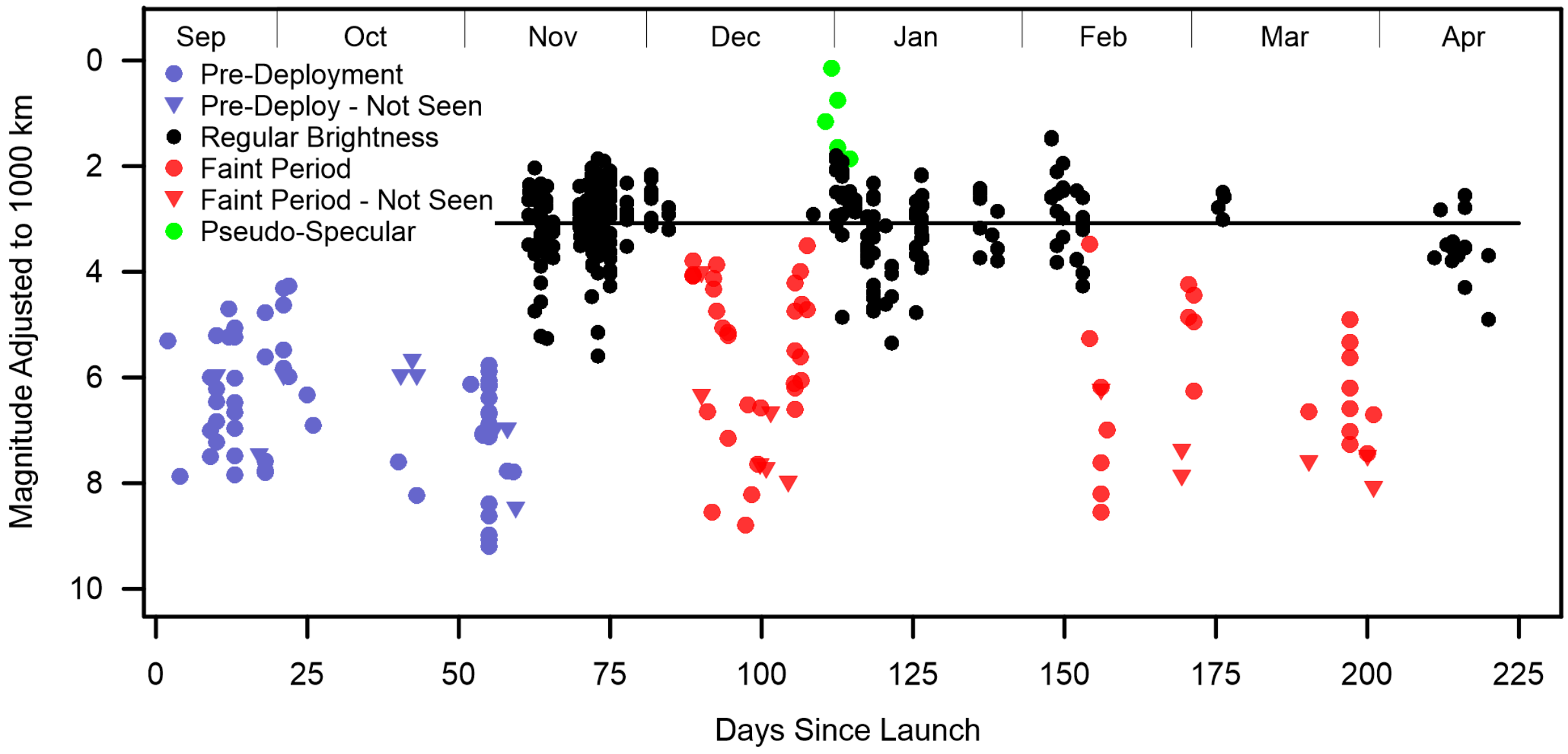

*Figure 1. The light curve of BW3 shows how brightness has changed since launch. Deployment around 2022 November 11 caused the satellite to become much more luminous (regular brightness). That was followed by faint periods in 2022 December and 2023 February and March. The symbols for 'not seen' indicate the magnitude of the faintest visible comparison star; so the satellite was dimmer than that value. The 'pseudo-specular' values are discussed in Section 4. The horizontal black line is the average of all regular magnitudes.*

The angle between the direction to the Sun and the plane of the satellite orbit is called beta (Versteeg and Cotton, and Sumanth 2019). The low power conditions described above occur when the cosine of the beta angle is a small value. Figure 2 offers evidence for the fading hypothesis described above by plotting the brightness of BW3 together with the cosine of its beta angle. The paper explaining the hypothesis was written while BW3 was still in its first faint period. This graph shows three complete cycles of fading and recovery.

Another way to visualize the brightness-beta relationship is shown in Figure 3 where magnitude is plotted as a function of cosine beta. The dividing line between regular and faint magnitudes occurs near cosine beta equals 0.8.

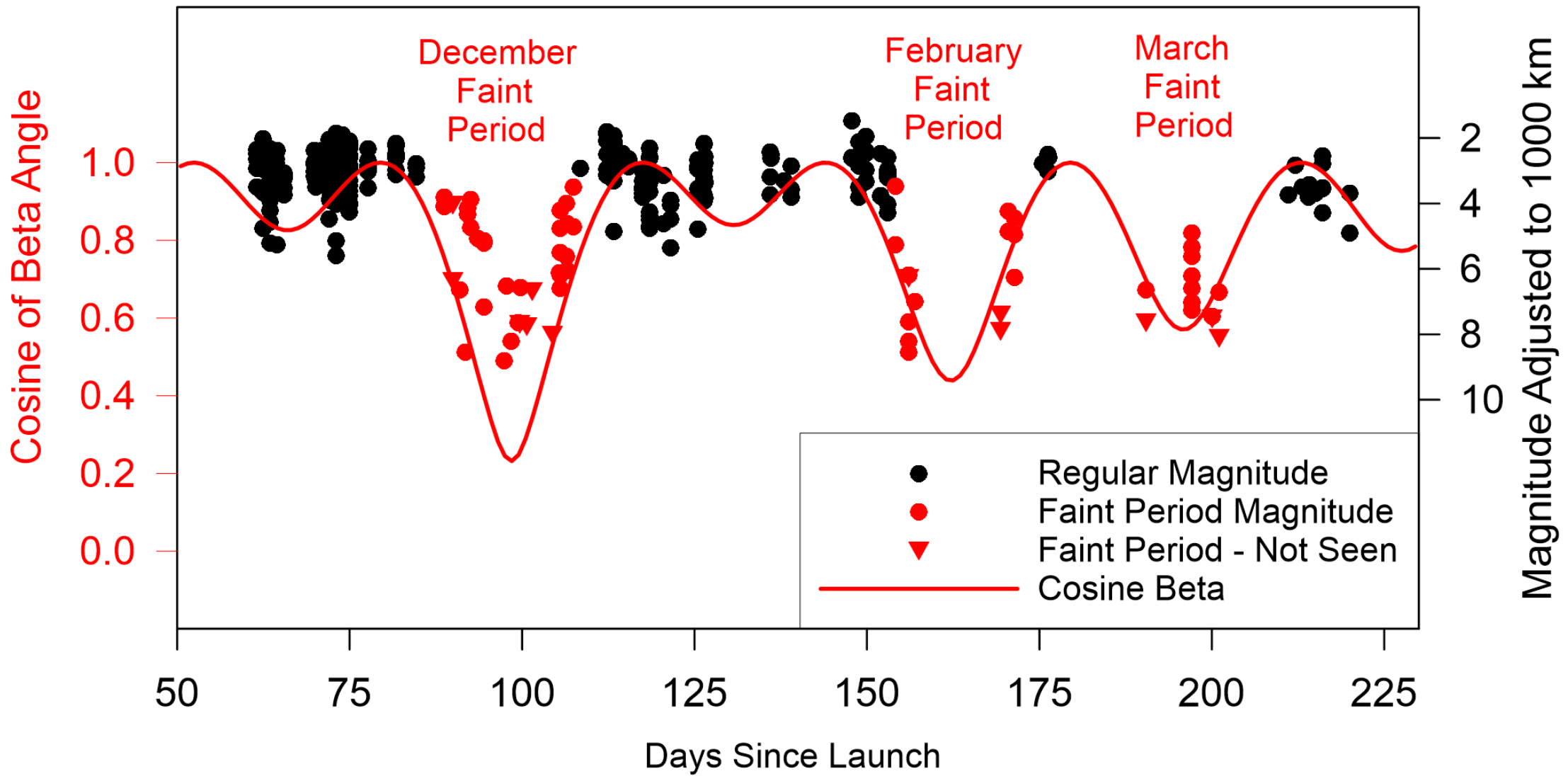

*Figure 2. All three brightness minima of BW3 coincide with minima of the cosine of the beta angle. The spacecraft solar array would have received very little insolation unless the zenith side of the flat panel was tilted toward the Sun.*

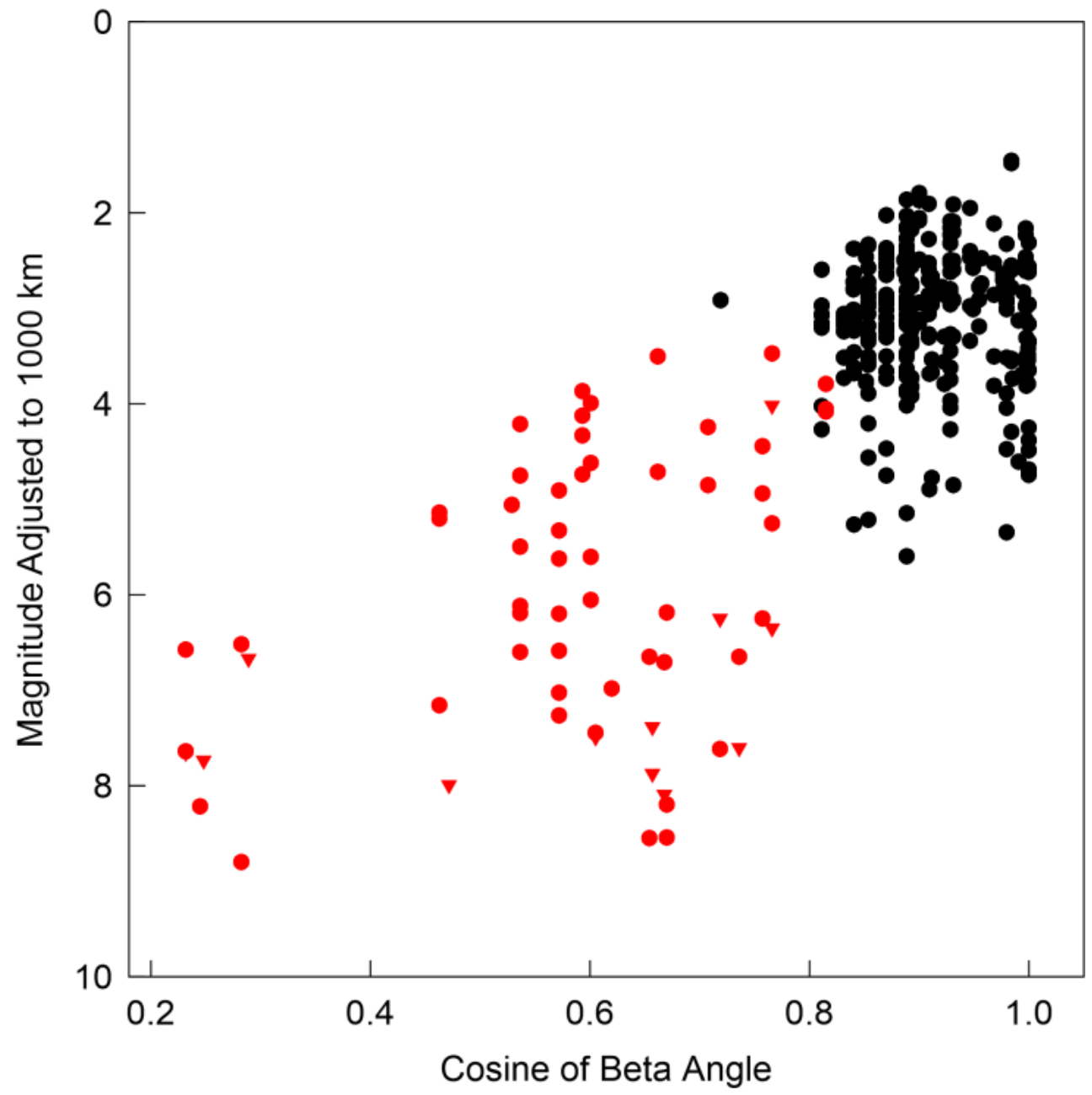

*Figure 3. Adjusted magnitude is plotted as a function of the cosine of the beta angle. Colors indicate regular and faint magnitudes as in Figure 2.*

3. Brightness characterization

This section addresses the distribution of apparent magnitudes, the illumination phase function and the peak brightness of BW3 when it is near zenith. The regular magnitudes analyzed here were obtained after deployment, outside of the faint periods and they do not include pseudo-specular observations.

The apparent brightness of a satellite is a direct measure of interference with astronomical observations. Magnitudes brighter than 7 are generally considered to problematic for astronomy (see the discussion in Tyson et al. 2020). The distribution of apparent visual magnitudes for BW3 is most often between 2.0 and 3.0 as illustrated in Figure 4. Magnitudes between 1.0 and 4.0 comprise 83% of the total. The average apparent magnitude is 2.75 with a standard deviation of 1.11 and a standard deviation of the mean of 0.07. The corresponding values for magnitudes adjusted to a standard distance of 1000 km are 3.09, 0.69 and 0.04. The smaller standard deviations for the distance-adjusted magnitudes indicate that range accounts for some of the brightness variation.

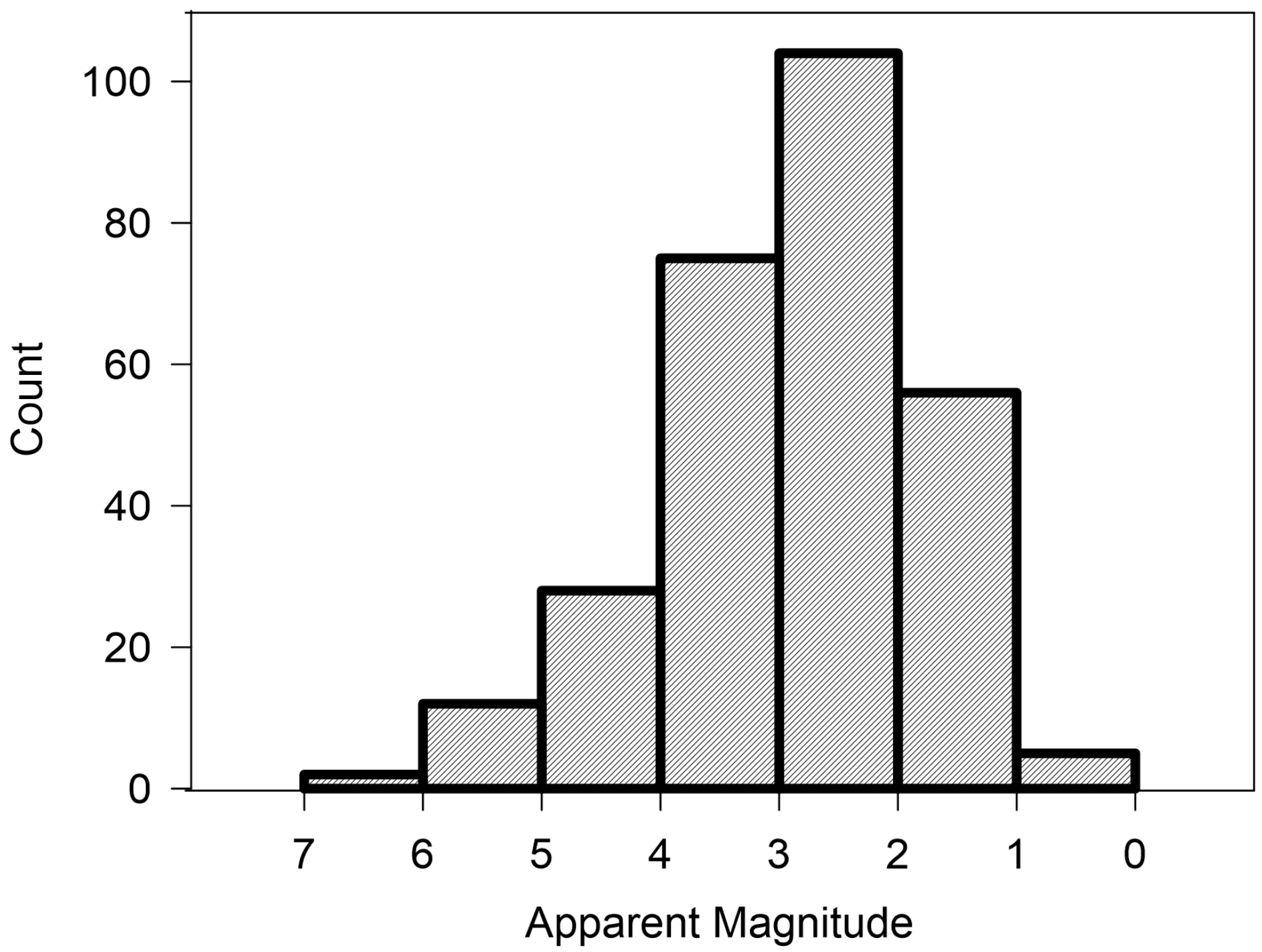

*Figure 4. The distribution of apparent visual magnitudes for BW3. Observations obtained before deployment and during the faint periods are excluded.*

The angle between the Sun and the observer measured at the satellite is known as its *phase angle*. The illumination phase function is the relationship between magnitude adjusted to a standard distance and phase angle. The phase function for BW3, shown in Figure 5, indicates that the satellite reflects sunlight most strongly at small to mid-range angles.

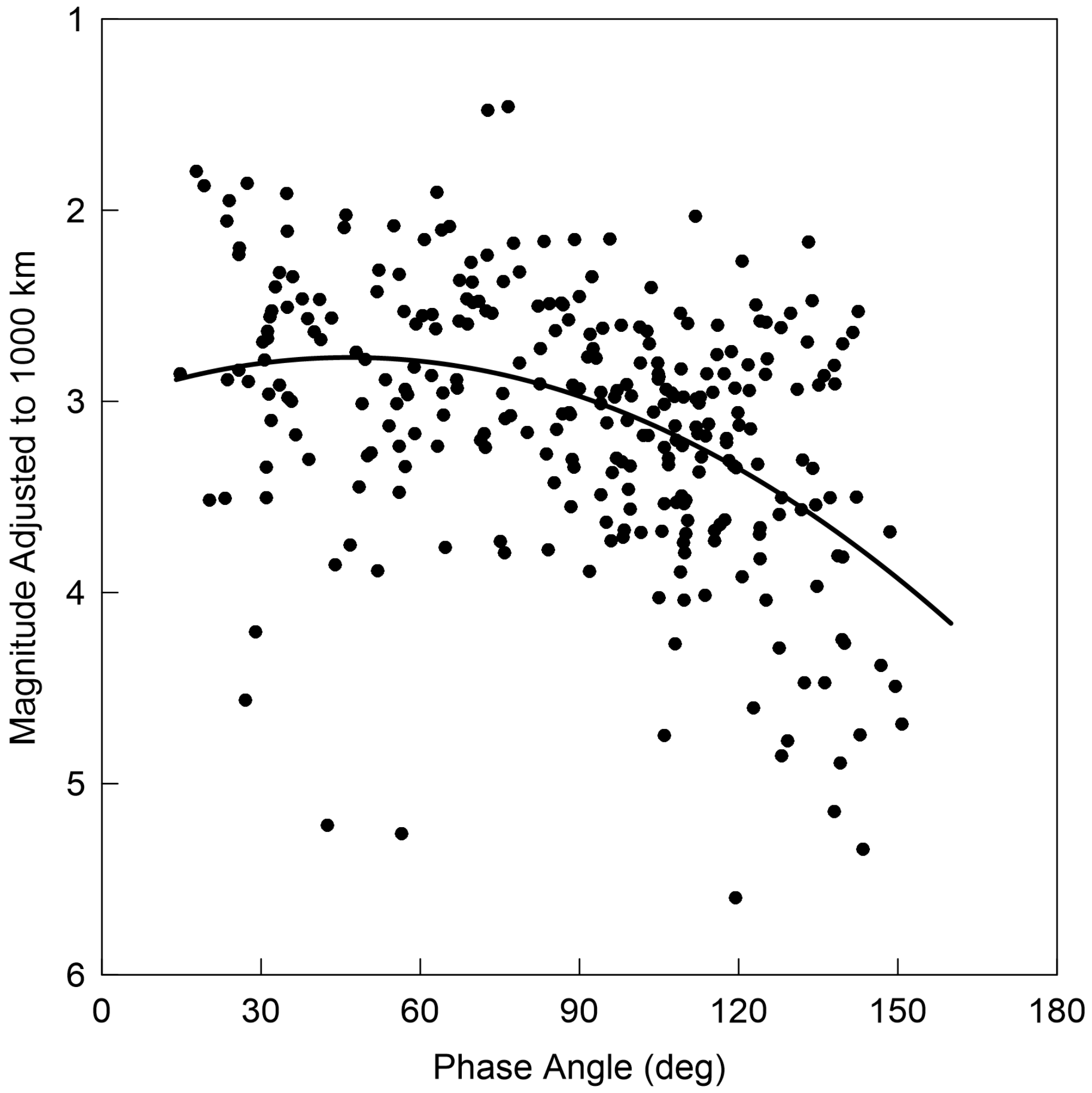

*Figure 5. The illumination phase function is brightest at phase angles less than 90°.*

The average brightness when a satellite is seen near zenith at the beginning and ending of astronomical twilight is termed the *characteristic magnitude by* Mallama et al. (2022). They described the method for deriving that value from the illumination phase curve and the distance to the satellite, and found that that characteristic magnitude for BW3 is 1.4. This compares to 4.7, 6.2 and 5.5 for Original, VisorSat and Post-VisorSat models of Starlink satellites (Mallama and Respler 2022). The addition of many new regular magnitudes since that 2022 analysis has not changed the characteristic magnitude of BW3.

Another way to evaluate the brightness of a satellite near zenith is to simply average the apparent magnitudes obtained at high elevations. The average magnitude for the BW3 magnitudes recorded above 60° elevation is also 1.4.

The analyses in this section described the brightness of BW3 empirically. The next section examines the luminosity in a physical sense using a model that accounts for the shape of the satellite, the reflective properties of its surface as well as its orientation in space.

4. A numerical model of the brightness of the BW3 spacecraft

Figure 6 shows the BW3 spacecraft before launch. The antenna panel which can be seen from the ground is visible. The solar panels are on the other face, but that face cannot be seen from the ground. The antenna panel appears in the image as a diffusely reflecting surface. There are no significant areas of specularly reflecting materials as are present on the Starlink spacecraft.

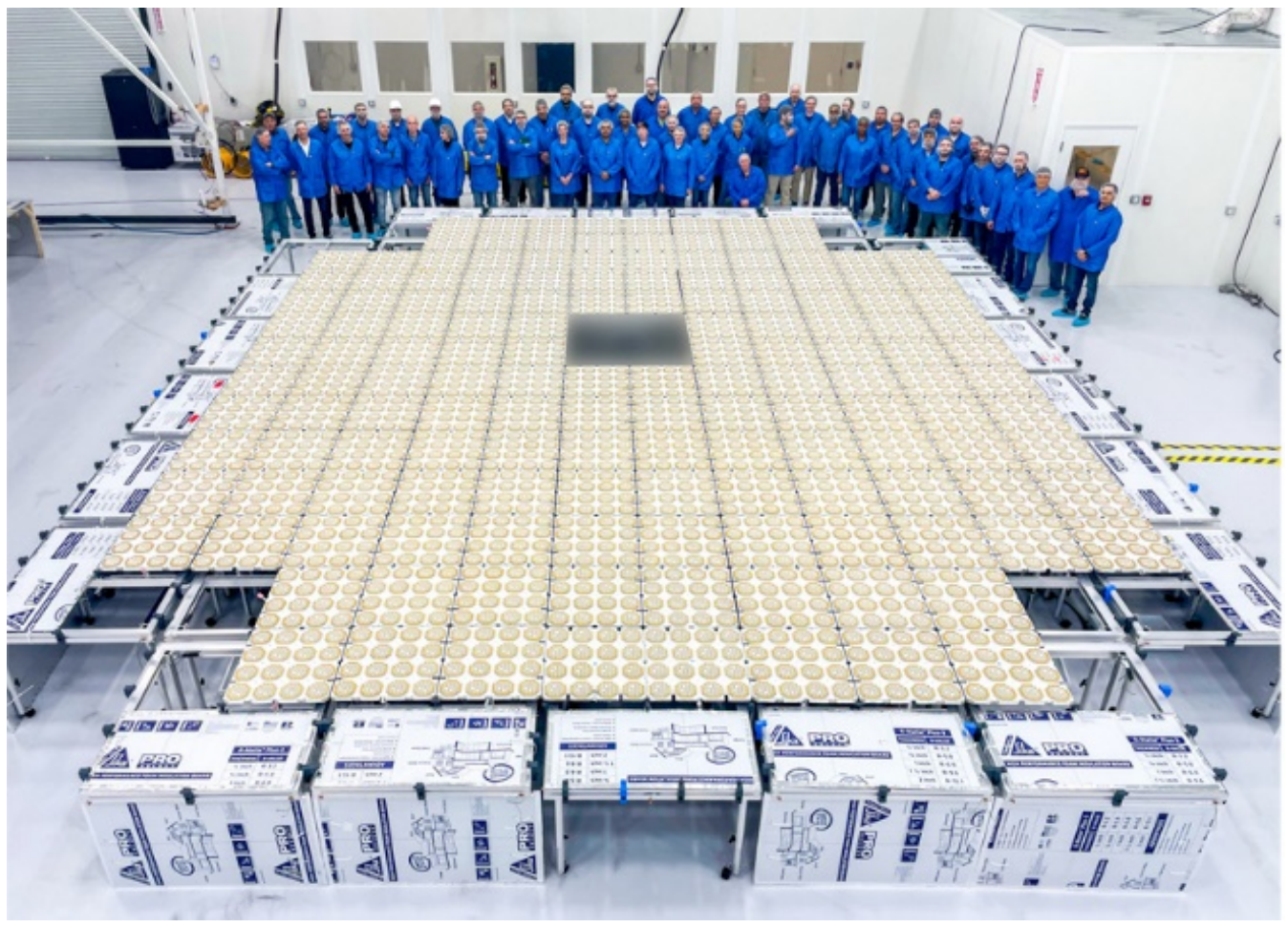

*Figure 6. The nadir face of the BlueWalker spacecraft during a pre-flight test at Cape Canaveral. The general appearance is of a diffusely reflecting flat panel (photo courtesy AST).*

A numerical model for the brightness of BW3 was developed, building on experience from observing and modeling Starlink spacecraft since 2020 (Cole 2020, 2021).

In the case of BW3, the numerical model uses a single Earth-facing flat surface that reflects light diffusely. Diffuse reflection may be represented by Lambert's Cosine Law, but this is not adequate if the line-of-sight is close to a specular reflected beam from the Sun. In these cases, BW3 can be a magnitude brighter than that predicted using Lambert's Law. This is because most diffusely reflecting materials will have a component that behaves in a near specular way. Simulating this component (termed here 'pseudo-specular' or PS component) is a standard feature of CGI – without it objects do not appear as in real-life. The PS component is significant when the angle of incidence is high (Figure 7), as is the case for all the observations of BW3 where the sunlight grazing angle on the nadir panel is less than 20°. The resulting model is termed a bidirectional reflectance distribution function (BRDF).

Earp et al (2007) provides a useful numerical model of the PS component that can be adapted for situations where the characteristics of the surface are not fully known. This was used to generate a BRDF that could be applied to the BW3. Example outputs are shown in Figure 8. The PS component becomes stronger as the angle of incidence approaches 90°, at the expense of the Lambertian component.

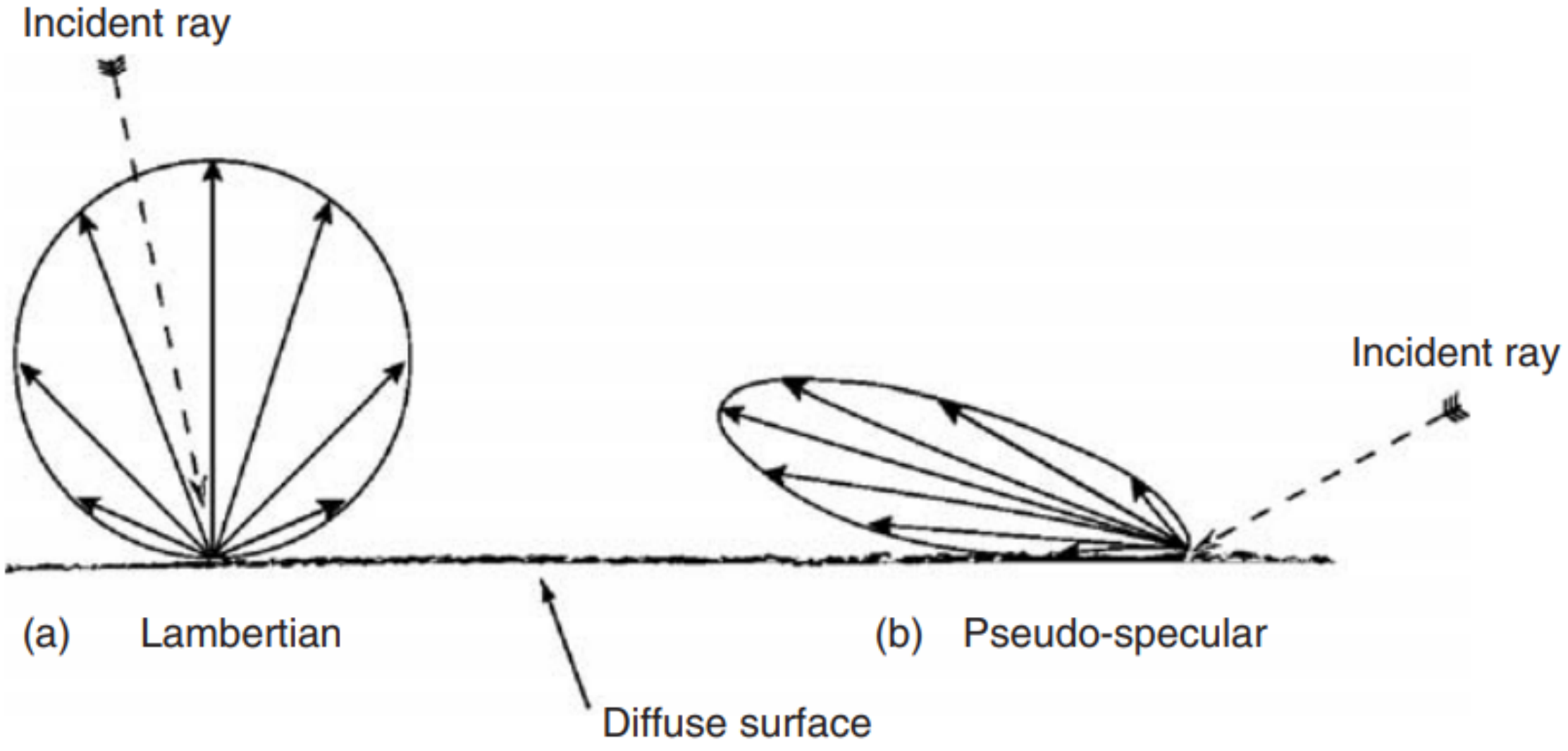

*Figure 7. Variation of reflectance profile with angle: (a) Lambertian reflection at small angles of incidence and (b) pseudo-specular reflection at large angles of incidence (from Earp et al, 2007).*

The model takes account of the aspect angle of the panel with respect to the observer, its range and the angle of the Sun illumination on the panel (which is different from the angle of the Sun at the observer).

The model was developed using observations made after full deployment of BW3. This dataset excludes the periods of dimming due to the tilt of the BW3 panel to the vertical. Data from those periods are discussed in Mallama et al (2023). A tilt angle was introduced into the model and was able to consistently match the observed BW3 brightness in those periods.

Outside the dimming periods, the panel was maintained in the model as nadir-facing and the only variables were the absolute brightness of the panel and the width of the pseudo-specular beam. With suitable selection of these two parameters, a good match was found between the predictions of the model and the visual observations (Figure 9). The BRDF improved the fit of a number of observations, particularly the fainter observations at the lower right of Figure 9. These were made at azimuths close to the Sun and at low elevations where the effect of the pseudo-specular beam (Figure 8) was visible. This occurs if the observation is within 30° of the specular reflected beam centerline.

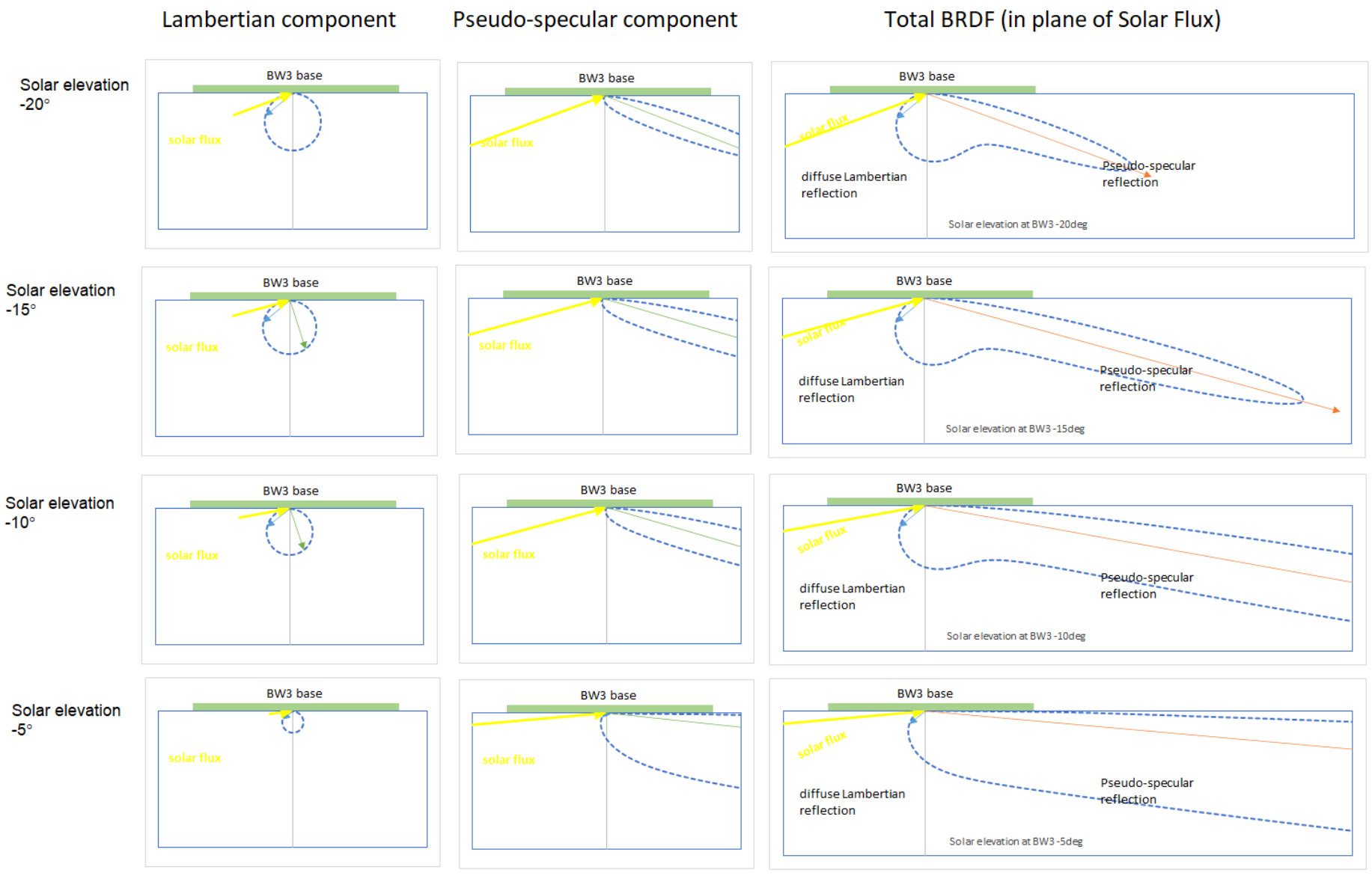

*Figure 8. The bidirectional reflectance distribution function (BRDF) as applied in modeling the reflectivity of BW3.*

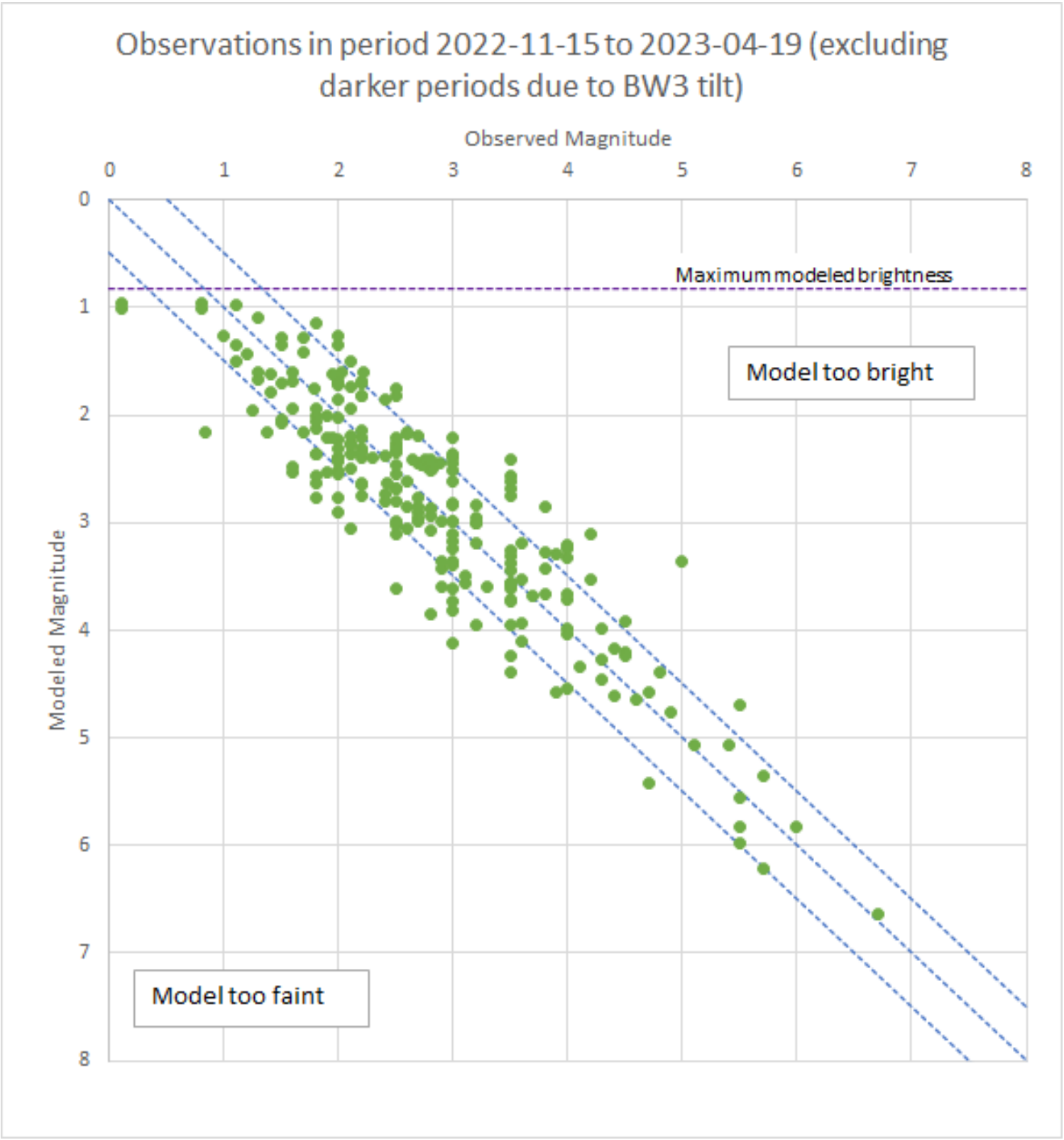

*Figure 9. Comparison of the brightness predictions of the numerical model and observations made in the period after the AST announcement of full deployment to April 2023. The closer the points are to the diagonal line, the better fit the model is to the observation.*

Using this model, brightness maps were created for the whole sky. Individual maps are required for each elevation of the Sun at the observer as this affects the appearance of the satellite across the whole sky. Figure 10 displays polar projection maps for Sun elevations of -18° (end of astronomical twilight) and -12° (end of nautical twilight). The following can be noted:

1. BW3 is in eclipse over part of the sky opposite the Sun – the anti-Sun direction (the white areas of the map)
2. The maximum brightness of BW3 is always in the zenith and is a strong function of the elevation of the Sun, brighter when the Sun is further below the horizon at the observer. BW3 is brightest when in the zenith and entering or emerging from eclipse.
3. The effect of the PS BRDF model is to increase the predicted brightness close the Sun azimuth and at low elevations, the brightness contours of the map are pulled to lower elevations in that direction.

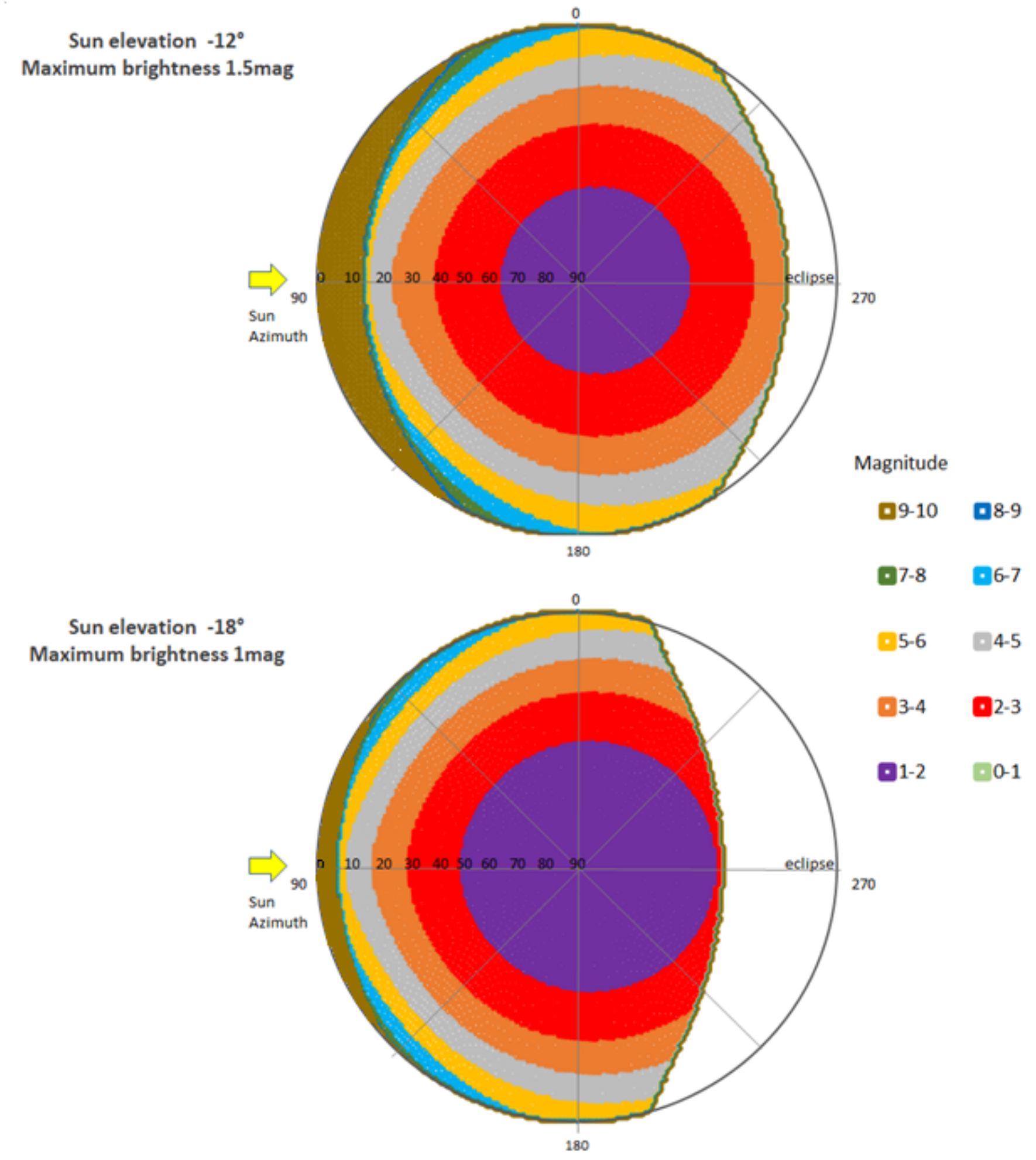

*Figure 10: Skymaps of BW3 brightness for sun elevations of -18° and -12°. A polar projection is used, north at the top and west at the right. The Sun azimuth of 90° is shown here but the relative appearance is the same for any Sun azimuth.*

5. Conclusions and discussion

The light curve of BW3 indicates that the folded spacecraft was relatively faint when it was launched. Then it became much brighter when it unfolded into a large flat-panel shape. This was followed by three separate intervals when the spacecraft faded and subsequently returned to its regular luminosity.

The distribution of apparent visual magnitudes for BW3 at its regular brightness is most often between 2.0 and 3.0. The characteristic brightness when seen near zenith is magnitude 1.4. The illumination phase function for BW3 indicates that it is brightest at small to mid-range phase angles.

A brightness model that represents the shape and orientation of the spacecraft and includes a reflection function with Lambertian and pseudo-specular components is applied to the observations. The analysis and modeling suggest that the zenith side of the flat-panel was tilted toward the Sun during the faint periods. We postulate that this adjustment was made in order to increase insolation on the solar array during an interval of low power. As a consequence, the nadir side, which faces observers on the ground, was only faintly lit or not lit at all by the Sun in those periods.

There are more intervals when BW3 is expected to fade during the coming months. Figure 11 plots cosine of the beta angle as in Figure 2 but it extends until 2023 August 26. There are deep minima in May and August along with other more modest dips.

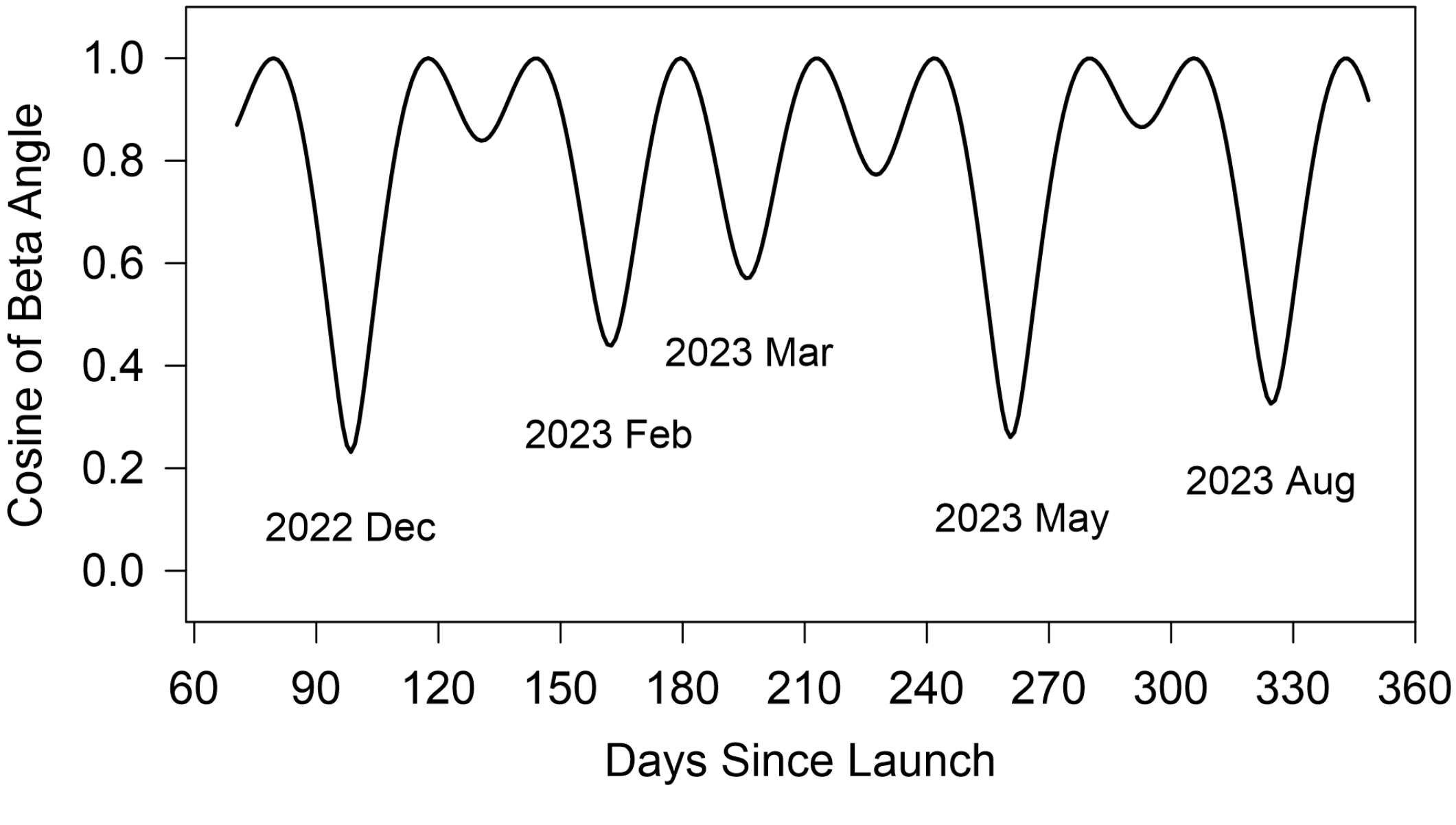

*Figure 11. The cosine of the beta angle is plotted through 2023 August. BW3 faded during the cosine minima of 2022 December and 2023 February and March. So, it may become dim again at such times in the future.*

Finally, the fading phenomenon appears to have arisen for operational reasons and it demonstrates that a small change of spacecraft attitude has a significant effect on the apparent magnitude of BW3. This finding can be used by satellite operators to purposefully make the satellite dimmer. That would reduce its interference with astronomical observations.

Acknowledgement

Several of the observers use the Heavens-Above.com web site maintained by Chris Peat to predict satellite passes.

References

Cole, R.E. 2020. A sky brightness model for the Starlink 'Visorsat' spacecraft – I, Research Notes of the American Astronomical Society, 4, 10, https://iopscience.iop.org/article/10.3847/2515-5172/abc0e9

Cole, R.E. 2021. A sky brightness model for the Starlink 'Visorsat' spacecraft. https://arxiv.org/abs/2107.06026

Earp, A., Smith,G. and Franklin, J. 2007. Simplified BRDF of a non-Lambertian diffuse surface, lighting. Res. Technol. 39, 3, pp. 265–281.

Halferty, G., Reddy, V., Campbell, T., Battle, A. and Furaro, R. 2022. Photometric characterization and trajectory accuracy of Starlink satellites: implications for ground-based astronomical surveys. https://arxiv.org/abs/2208.03226.

Mallama, A., 2022. The method of visual satellite photometry. https://arxiv.org/abs/2208.07834.

Mallama, A. and Respler, J. 2022. Visual brightness characteristics of Starlink Generation 1 satellites. https://arxiv.org/abs/2210.17268.

Mallama, A., Cole, R.E., Harrington, S. and Maley, P.D. 2022. Visual magnitude of the BlueWalker 3 satellite. https://arxiv.org/abs/2211.09811.

Mallama, A., Cole, R.E. and Tilley, S. 2023. The BlueWalker 3 satellite has faded. https://arxiv.org/abs/2301.01601

Mroz, P., Otarola, A., Prince, T.A., Dekany, R., Duev, D.A., Graham, M.J., Groom, S.L., Masci, F.J. and Medford, M.S. 2022. Impact of the SpaceX Starlink satellites on the Zwicky Transient Facility survey observations. https://arxiv.org/abs/2201.05343.

Sumanth, R.M. 2019. Computation of eclipse time for low-Earth orbiting small satellites. https://commons.erau.edu/cgi/viewcontent.cgi?article=1412&context=ijaaa

Tyson, J. A. and 10 co-authors. 2020. Mitigation of LEO satellite brightness and trail effects on the Rubin Observatory LSST. Astron. J. 160, 226 and https://arxiv.org/abs/2006.12417.

Versteeg, C. and Cotten, D.L. Preliminary thermal analysis of small satellites. https://s3vi.ndc.nasa.gov/ssri-kb/static/resources/Preliminary_Thermal_Analysis_of_Small_Satellites.pdf

Walker, C. and Benvenuti, P. (Eds.) 2022. Dark and Quite Skies for Science and Society II. Working group reports. https://noirlab.edu/public/media/archives/techdocs/pdf/techdoc051.pdf.

## Appendix A. Observer coordinates

| Observer | Latitude | Longitude | Ht(m) | Site |
|---|---|---|---|---|
| J. Barentine | 32.234 | -110.768 | 833 | |
| C. Bassa | 52.909 | 6.869 | | |
| P. Creed | 40.893 | -81.536 | | |
| R. Cole | 50.552 | -4.735 | 100 | |
| K. Fetter | 44.606 | -75.691 | | |
| S. Harrington | 36.062 | -91.688 | 185 | |
| M. Langbroek | 52.154 | 4.491 | 0 | |
| R. Lee | 38.93 | -104.81 | 2082 | |
| P. Maley | 33.811 | -111.952 | 654 | 1 |
| P. Maley | 32.857 | -113.220 | | 2 |
| P. Maley | 34.6 | 33.0 | 0 | 3 |
| A. Mallama | 38.982 | -76.763 | 43 | 1 |
| A. Mallama | 38.72 | -75.08 | 0 | 2 |
| A. Mallama | 39.122 | -77.891 | | 3 |
| R. McNaught | -32.27 | 149.16 | 610 | |
| J. Respler | 40.330 | -74.445 | | |
| R. Swaney | 41.403 | -81.512 | | |
| S. Tilley | 49.434 | -123.668 | 40 | 1 |
| S. Tilley | 49.418 | -123.642 | 1 | 2 |
| E. Visser | 53.109 | 6.108 | 46 | |
| A. Worley | 41.474 | -81.519 | 351 | |
| J. Worley | 41.474 | -81.519 | 351 | |
| B. Young | 36.139 | -95.983 | 201 | 1 |
| B. Young | 35.831 | -96.141 | 330 | 2 |

## Appendix B. Observations

These observations have sufficient geometrical information for all the analyses described in this paper. Some additional observations with only apparent magnitudes were used in earlier studies. The symbol '>' indicates the magnitude of the faintest comparison star observed when the satellites was too faint to be seen. Site numbers apply where observers recorded data from more than one location. The default site number is '1' and all coordinates are listed in Appendix A. Type is visual unless otherwise indicated.

| Observer | UTC Date | UTC Time | Mag. | Site | Type |
|---|---|---|---|---|---|
| Maley | 2022-Sep-13 | 2:30:43 | 5.2 | | |
| Mallama | 2022-Sep-15 | 0:23:38 | 6.7 | | |
| Cole | 2022-Sep-20 | 19:22:00 | 6.1 | | |
| Cole | 2022-Sep-20 | 20:59:40 | 6.6 | | |
| Cole | 2022-Sep-20 | 21:00:45 | 5 | | |
| Mallama | 2022-Sep-21 | 0:04:13 | > 6 | | |
| Langbroek | 2022-Sep-21 | 19:04:57 | 6.25 | | Video |
| Langbroek | 2022-Sep-21 | 19:05:17 | 5.62 | | Video |
| Langbroek | 2022-Sep-21 | 19:05:28 | 5.12 | | Video |
| Cole | 2022-Sep-21 | 20:42:00 | 5.7 | | |
| Cole | 2022-Sep-21 | 20:43:00 | 4.2 | | |
| Cole | 2022-Sep-23 | 20:05:55 | 4.1 | | |
| Cole | 2022-Sep-23 | 20:06:54 | 4.2 | | |
| Langbroek | 2022-Sep-24 | 19:49:00 | 7.02 | | Video |
| Langbroek | 2022-Sep-24 | 19:49:40 | 5.60 | | Video |
| Langbroek | 2022-Sep-24 | 19:50:15 | 3.85 | | Video |
| Mallama | 2022-Sep-28 | 1:17:45 | > 7.5 | | |
| Cole | 2022-Sep-29 | 19:57:08 | 4.55 | | |
| Cole | 2022-Sep-29 | 19:57:35 | 3.5 | | |
| Langbroek | 2022-Sep-29 | 19:58:09 | 7.55 | | Video |
| Langbroek | 2022-Sep-29 | 19:58:13 | 7.56 | | Video |
| Langbroek | 2022-Sep-29 | 19:58:30 | 7.27 | | Video |
| Tilley | 2022-Oct-02 | 3:15:52 | 3.0 | | Video |
| Maley | 2022-Oct-02 | 3:18:00 | > 6.0 | | |
| Cole | 2022-Oct-02 | 19:02:23 | 5.3 | | |
| Cole | 2022-Oct-02 | 19:03:07 | 3.6 | | |
| Cole | 2022-Oct-02 | 19:04:02 | 4.4 | | |
| Tilley | 2022-Oct-03 | 2:57:49 | 3.0 | | Video |
| Tilley | 2022-Oct-03 | 2:58:40 | 5.0 | | Video |
| Maley | 2022-Oct-06 | 2:09:10 | 6.0 | | |
| Mallama | 2022-Oct-07 | 0:17:15 | 6.9 | 2 | |
| Harrington | 2022-Oct_21 | 11:22:00 | 7.4 | | |
| Maley | 2022-Oct_21 | 12:57:00 | > 6.0 | | |
| Cole | 2022-Oct_23 | 6:06:00 | > 5.7 | | |
| Maley | 2022-Oct_24 | 2:36:00 | > 6.0 | 2 | |
| Cole | 2022-Oct_24 | 5:48:00 | 8.5 | | |
| Cole | 2022-Nov-02 | 6:22:39 | 5.5 | | |
| Cole | 2022-Nov-04 | 5:46:19 | 6.4 | | |
| Cole | 2022-Nov-04 | 5:46:37 | 6.1 | | |
| Bassa | 2022-Nov-05 | 5:28:08 | 6.3 | | CMOS |
| Bassa | 2022-Nov-05 | 5:28:18 | 6.3 | | CMOS |

| | | | | |
|---|---|---|---|---|
| Bassa | 2022-Nov-05 | 5:28:40 | 6.3 | CMOS |
| Bassa | 2022-Nov-05 | 5:28:48 | 6.2 | CMOS |
| Bassa | 2022-Nov-05 | 5:28:58 | 6.0 | CMOS |
| Bassa | 2022-Nov-05 | 5:29:08 | 6.0 | CMOS |
| Bassa | 2022-Nov-05 | 5:29:17 | 6.1 | CMOS |
| Bassa | 2022-Nov-05 | 5:29:38 | 6.1 | CMOS |
| Bassa | 2022-Nov-05 | 5:29:48 | 6.0 | CMOS |
| Bassa | 2022-Nov-05 | 5:29:55 | 6.1 | CMOS |
| Bassa | 2022-Nov-05 | 5:30:09 | 6.2 | CMOS |
| Bassa | 2022-Nov-05 | 5:30:17 | 6.1 | CMOS |
| Bassa | 2022-Nov-05 | 5:31:08 | 7.3 | CMOS |
| Bassa | 2022-Nov-05 | 5:31:18 | 7.6 | CMOS |
| Bassa | 2022-Nov-05 | 5:31:41 | 8.2 | CMOS |
| Bassa | 2022-Nov-05 | 5:31:48 | 8.5 | CMOS |
| Bassa | 2022-Nov-05 | 5:31:58 | 8.5 | CMOS |
| Mallama | 2022-Nov-08 | 10:58:25 | 7.2 | |
| Mallama | 2022-Nov-08 | 10:59:25 | > 7.0 | |
| McNaught | 2022-Nov-09 | 10:00:00 | > 8.5 | |
| Mallama | 2022-Nov-09 | 10:40:15 | 7.1 | |
| Harrington | 2022-Nov-11 | 11:40:00 | 2.0 | |
| Harrington | 2022-Nov-11 | 11:41:00 | 1.5 | |
| Harrington | 2022-Nov-11 | 11:41:40 | 2.1 | |
| Maley | 2022-Nov-11 | 13:15:37 | 3.5 | |
| Maley | 2022-Nov-11 | 13:16:57 | 1.0 | |
| Harrington | 2022-Nov-12 | 11:21:10 | 2.5 | |
| Harrington | 2022-Nov-12 | 11:22:00 | 1.7 | |
| Harrington | 2022-Nov-12 | 11:23:00 | 1.0 | |
| Harrington | 2022-Nov-12 | 11:23:30 | 1.6 | |
| Harrington | 2022-Nov-12 | 11:24:00 | 2.0 | |
| Harrington | 2022-Nov-12 | 11:25:00 | 3.6 | |
| Maley | 2022-Nov-12 | 12:57:00 | 3.5 | |
| Maley | 2022-Nov-12 | 12:58:00 | 1.0 | |
| Maley | 2022-Nov-12 | 12:59:00 | 3.5 | |
| Harrington | 2022-Nov-13 | 11:04:12 | 4.0 | |
| Harrington | 2022-Nov-13 | 11:04:46 | 0.9 | |
| Harrington | 2022-Nov-13 | 11:06:00 | 2.1 | |
| Harrington | 2022-Nov-13 | 11:07:00 | 3.2 | |
| Harrington | 2022-Nov-13 | 11:08:00 | 4.4 | |
| Maley | 2022-Nov-13 | 12:39:11 | 4.0 | |
| Maley | 2022-Nov-13 | 12:39:20 | 3.5 | |
| Maley | 2022-Nov-13 | 12:39:30 | 2.5 | |
| Maley | 2022-Nov-13 | 12:39:39 | 2.0 | |
| Maley | 2022-Nov-13 | 12:39:48 | 2.2 | |
| Maley | 2022-Nov-13 | 12:40:02 | 2.0 | |
| Maley | 2022-Nov-13 | 12:40:06 | 1.0 | |
| Maley | 2022-Nov-13 | 12:41:23 | 1.5 | |
| Maley | 2022-Nov-13 | 12:41:32 | 2.0 | |
| Maley | 2022-Nov-13 | 12:41:41 | 2.5 | |
| Maley | 2022-Nov-13 | 12:41:59 | 3.0 | |
| Maley | 2022-Nov-13 | 12:42:09 | 3.5 | |
| Harrington | 2022-Nov-14 | 10:47:30 | 1.9 | |
| Harrington | 2022-Nov-14 | 10:48:00 | 2.2 | |
| Harrington | 2022-Nov-14 | 10:49:00 | 3.1 | |
| Maley | 2022-Nov-14 | 12:22:22 | 4.0 | |
| Maley | 2022-Nov-14 | 12:22:32 | 2.0 | |
| Maley | 2022-Nov-14 | 12:22:42 | 1.2 | |
| Maley | 2022-Nov-14 | 12:22:52 | 2.0 | |
| Maley | 2022-Nov-14 | 12:23:12 | 2.2 | |
| Maley | 2022-Nov-14 | 12:23:32 | 2.5 | |
| Maley | 2022-Nov-14 | 12:23:42 | 3.0 | |
| Maley | 2022-Nov-14 | 12:24:02 | 3.5 | |
| Maley | 2022-Nov-14 | 12:24:22 | 3.7 | |
| Maley | 2022-Nov-15 | 12:05:32 | 3.5 | |
| Maley | 2022-Nov-15 | 12:05:42 | 3.5 | |
| Maley | 2022-Nov-15 | 12:05:52 | 3.0 | |
| Maley | 2022-Nov-15 | 12:06:02 | 3.2 | |
| Maley | 2022-Nov-15 | 12:06:12 | 3.2 | |
| Maley | 2022-Nov-15 | 12:06:22 | 3.5 | |
| Maley | 2022-Nov-15 | 12:06:32 | 3.5 | |
| Maley | 2022-Nov-15 | 12:06:42 | 4.0 | |
| Mallama | 2022-Nov-19 | 23:01:11 | 4.0 | |
| Mallama | 2022-Nov-19 | 23:01:22 | 3.5 | |
| Mallama | 2022-Nov-19 | 23:01:37 | 3.5 | |
| Mallama | 2022-Nov-19 | 23:01:50 | 3.0 | |
| Mallama | 2022-Nov-19 | 23:02:55 | 3.5 | |
| Mallama | 2022-Nov-19 | 23:03:14 | 3.5 | |
| Harrington | 2022-Nov-20 | 0:37:00 | 2.7 | |
| Harrington | 2022-Nov-20 | 0:38:30 | 1.7 | |
| Maley | 2022-Nov-20 | 2:10:31 | 4.0 | 3 |
| Maley | 2022-Nov-20 | 2:10:59 | 3.5 | 3 |
| Maley | 2022-Nov-20 | 2:11:09 | 3.0 | 3 |
| Maley | 2022-Nov-20 | 2:11:45 | 2.7 | 3 |
| Maley | 2022-Nov-20 | 2:12:11 | 2.0 | 3 |
| Maley | 2022-Nov-20 | 2:12:38 | 1.5 | 3 |
| Respler | 2022-Nov-20 | 22:45:19 | 3.6 | |
| Harrington | 2022-Nov-21 | 0:18:30 | 2.9 | |

| | | | |
|---|---|---|---|
| Harrington | 2022-Nov-21 | 0:20:15 | 1.1 |
| Harrington | 2022-Nov-21 | 0:21:15 | 2.2 |
| Swaney | 2022-Nov-21 | 0:21:20 | 2.8 |
| Maley | 2022-Nov-21 | 1:53:04 | 3.5 |
| Maley | 2022-Nov-21 | 1:53:14 | 3.0 |
| Maley | 2022-Nov-21 | 1:53:49 | 2.5 |
| Maley | 2022-Nov-21 | 1:54:40 | 2.0 |
| Maley | 2022-Nov-21 | 1:55:06 | 1.5 |
| Maley | 2022-Nov-21 | 1:55:41 | 2.0 |
| Maley | 2022-Nov-21 | 1:56:19 | 2.5 |
| Harrington | 2022-Nov-21 | 23:58:30 | 5.5 |
| Harrington | 2022-Nov-21 | 0:00:30 | 2.9 |
| Harrington | 2022-Nov-21 | 0:01:00 | 2.5 |
| Harrington | 2022-Nov-21 | 0:02:15 | 1.2 |
| Harrington | 2022-Nov-21 | 0:03:30 | 2.4 |
| Harrington | 2022-Nov-21 | 0:04:00 | 2.8 |
| Mallama | 2022-Nov-22 | 0:03:50 | 2.0 |
| Mallama | 2022-Nov-22 | 0:04:14 | 1.5 |
| Swaney | 2022-Nov-22 | 0:04:20 | 0.8 |
| Maley | 2022-Nov-22 | 1:34:55 | 3.0 |
| Maley | 2022-Nov-22 | 1:35:28 | 2.5 |
| Maley | 2022-Nov-22 | 1:35:59 | 2.0 |
| Maley | 2022-Nov-22 | 1:36:21 | 2.2 |
| Maley | 2022-Nov-22 | 1:36:52 | 2.5 |
| Maley | 2022-Nov-22 | 1:37:02 | 2.2 |
| Maley | 2022-Nov-22 | 1:37:23 | 2.0 |
| Maley | 2022-Nov-22 | 1:38:11 | 2.5 |
| Maley | 2022-Nov-22 | 1:38:20 | 3.0 |
| Maley | 2022-Nov-22 | 1:38:31 | 3.5 |
| Maley | 2022-Nov-22 | 1:39:19 | 4.0 |
| Mallama | 2022-Nov-22 | 23:45:32 | 3.5 |
| Mallama | 2022-Nov-22 | 23:45:47 | 3.0 |
| Mallama | 2022-Nov-22 | 23:46:04 | 2.5 |
| Mallama | 2022-Nov-22 | 23:46:13 | 2.0 |
| Mallama | 2022-Nov-22 | 23:46:26 | 1.8 |
| Mallama | 2022-Nov-22 | 23:46:43 | 1.5 |
| Mallama | 2022-Nov-22 | 23:46:55 | 1.8 |
| Mallama | 2022-Nov-22 | 23:47:05 | 2.0 |
| Swaney | 2022-Nov-22 | 23:45:10 | 3.0 |
| Swaney | 2022-Nov-22 | 23:47:05 | 1.3 |
| Harrington | 2022-Nov-22 | 23:41:00 | 5.7 |
| Harrington | 2022-Nov-22 | 23:42:30 | 5.0 |
| Harrington | 2022-Nov-22 | 23:42:45 | 3.2 |

| | | | | |
|---|---|---|---|---|
| Harrington | 2022-Nov-22 | 23:43:15 | 2.5 | |
| Harrington | 2022-Nov-22 | 23:44:00 | 1.5 | |
| Harrington | 2022-Nov-22 | 23:45:00 | 2.3 | |
| Harrington | 2022-Nov-22 | 23:47:00 | 4.3 | |
| Worley | 2022-Nov-22 | 23:46:25 | 1.1 | |
| Mallama | 2022-Nov-23 | 23:25:56 | 3.0 | 3 |
| Mallama | 2022-Nov-23 | 23:26:13 | 3.0 | 3 |
| Mallama | 2022-Nov-23 | 23:26:34 | 3.0 | 3 |
| Mallama | 2022-Nov-23 | 23:27:25 | 2.5 | 3 |
| Mallama | 2022-Nov-23 | 23:27:46 | 2.0 | 3 |
| Mallama | 2022-Nov-23 | 23:27:58 | 2.2 | 3 |
| Mallama | 2022-Nov-23 | 23:28:12 | 1.9 | 3 |
| Mallama | 2022-Nov-23 | 23:28:30 | 2.6 | 3 |
| Mallama | 2022-Nov-23 | 23:28:40 | 2.6 | 3 |
| Mallama | 2022-Nov-23 | 23:29:05 | 2.0 | 3 |
| Mallama | 2022-Nov-23 | 23:29:12 | 1.8 | 3 |
| Mallama | 2022-Nov-23 | 23:29:30 | 1.6 | 3 |
| Respler | 2022-Nov-23 | 23:29:30 | 2.5 | |
| Cole | 2022-Nov-24 | 18:27:40 | 2.1 | |
| Mallama | 2022-Nov-24 | 23:08:06 | 4.5 | 3 |
| Mallama | 2022-Nov-24 | 23:08:26 | 4.0 | 3 |
| Mallama | 2022-Nov-24 | 23:08:56 | 3.8 | 3 |
| Mallama | 2022-Nov-24 | 23:09:16 | 3.2 | 3 |
| Mallama | 2022-Nov-24 | 23:09:26 | 2.8 | 3 |
| Mallama | 2022-Nov-24 | 23:09:38 | 2.4 | 3 |
| Mallama | 2022-Nov-24 | 23:09:50 | 2.2 | 3 |
| Mallama | 2022-Nov-24 | 23:10:04 | 2.0 | 3 |
| Mallama | 2022-Nov-24 | 23:10:19 | 1.9 | 3 |
| Mallama | 2022-Nov-24 | 23:10:36 | 1.8 | 3 |
| Mallama | 2022-Nov-24 | 23:10:50 | 1.8 | 3 |
| Mallama | 2022-Nov-24 | 23:11:08 | 1.8 | 3 |
| Mallama | 2022-Nov-24 | 23:11:22 | 2.0 | 3 |
| Mallama | 2022-Nov-24 | 23:11:35 | 2.5 | 3 |
| Mallama | 2022-Nov-24 | 23:11:49 | 3.0 | 3 |
| Mallama | 2022-Nov-24 | 23:12:01 | 3.5 | 3 |
| Mallama | 2022-Nov-24 | 23:12:13 | 3.8 | 3 |
| Mallama | 2022-Nov-24 | 23:12:23 | 4.2 | 3 |
| Cole | 2022-Nov-27 | 17:32:07 | 2.2 | |
| Cole | 2022-Nov-27 | 17:32:26 | 2.0 | |
| Cole | 2022-Nov-27 | 17:33:06 | 2.1 | |
| Cole | 2022-Nov-27 | 17:33:35 | 2.1 | |
| Cole | 2022-Nov-27 | 17:33:50 | 2.0 | |
| Cole | 2022-Nov-27 | 17:34:02 | 2.4 | |
| Cole | 2022-Nov-27 | 17:34:14 | 2.7 | |
| Cole | 2022-Nov-27 | 17:34:33 | 3.6 | |
| Cole | 2022-Dec-01 | 17:56:01 | 2.5 | |
| Cole | 2022-Dec-01 | 17:56:31 | 2.7 | |
| Cole | 2022-Dec-01 | 17:56:59 | 2.2 | |
| Cole | 2022-Dec-01 | 17:57:11 | 1.7 | |
| Cole | 2022-Dec-01 | 17:58:00 | 1.3 | |
| Cole | 2022-Dec-01 | 17:58:55 | 1.8 | |
| Cole | 2022-Dec-01 | 17:59:16 | 1.8 | |
| Cole | 2022-Dec-01 | 19:33:27 | 3.1 | |
| McNaught | 2022-Dec-04 | 16:07:23 | 3.7 | |
| McNaught | 2022-Dec-04 | 16:07:57 | 3.5 | |
| McNaught | 2022-Dec-04 | 16:09:12 | 4.1 | |
| Langbroek | 2022-Dec-08 | 17:30:55 | 3.81 | Video |
| Langbroek | 2022-Dec-08 | 17:31:10 | 3.95 | Video |
| Langbroek | 2022-Dec-08 | 17:31:14 | 4.01 | Video |
| Langbroek | 2022-Dec-08 | 17:31:17 | 3.82 | Video |
| Mallama | 2022-Dec-09 | 23:31:30 | > 7.0 | |
| Mallama | 2022-Dec-09 | 23:33:55 | > 3.0 | |
| Lee | 2022-Dec-11 | 0:49:25 | 6.0 | |
| Cole | 2022-Dec-11 | 18:10:13 | 8.2 | |
| Barentine | 2022-Dec-12 | 2:10:01 | 3.0 | |
| Barentine | 2022-Dec-12 | 2:10:28 | 3.0 | |
| McNaught | 2022-Dec-12 | 13:41:15 | 5.1 | |
| McNaught | 2022-Dec-12 | 13:41:47 | 6.1 | |
| McNaught | 2022-Dec-13 | 13:22:44 | 6.3 | |
| McNaught | 2022-Dec-14 | 11:27:17 | 8.5 | |
| McNaught | 2022-Dec-14 | 11:27:38 | 6.7 | |
| McNaught | 2022-Dec-14 | 13:04:12 | 6.4 | |
| McNaught | 2022-Dec-17 | 10:25:48 | 8.7 | |
| McNaught | 2022-Dec-17 | 17:09:30 | 7.4 | |
| McNaught | 2022-Dec-18 | 10:11:38 | 8.5 | |
| McNaught | 2022-Dec-19 | 9:50:43 | 6.4 | |
| McNaught | 2022-Dec-19 | 18:10:17 | 5.6 | |
| McNaught | 2022-Dec-19 | 16:30:55 | > 8.5 | |
| McNaught | 2022-Dec-20 | 16:12:23 | > 8.5 | |
| Mallama | 2022-Dec-21 | 11:36:05 | > 6.8 | |
| Mallama | 2022-Dec-24 | 10:39:50 | > 7.9 | |
| Mallama | 2022-Dec-25 | 10:21:50 | 5.8 | |
| Harrington | 2022-Dec-25 | 11:55:00 | 4.5 | |
| Harrington | 2022-Dec-25 | 11:56:00 | 4.3 | |
| Harrington | 2022-Dec-25 | 11:57:00 | 4.3 | |
| Harrington | 2022-Dec-25 | 11:57:30 | 4.8 | |

| | | | |
|---|---|---|---|
| Harrington | 2022-Dec-25 | 11:58:00 | 5.3 |
| Mallama | 2022-Dec-26 | 10:03:25 | 5.2 |
| Mallama | 2022-Dec-26 | 11:40:55 | 6.1 |
| Respler | 2022-Dec-26 | 10:03:50 | 3.5 |
| McNaught | 2022-Dec-26 | 14:19:26 | 5.8 |
| Harrington | 2022-Dec-27 | 11:22:00 | 3.1 |
| Harrington | 2022-Dec-27 | 11:23:00 | 5.0 |
| Mallama | 2022-Dec-28 | 11:04:50 | 2.8 |
| McNaught | 2022-Dec-30 | 13:05:30 | 2.1 |
| McNaught | 2022-Dec-31 | 12:46:42 | 1.2 |
| Cole | 2023-Jan-01 | 6:45:18 | 1.6 |
| Cole | 2023-Jan-01 | 6:45:24 | 1.6 |
| Cole | 2023-Jan-01 | 6:45:39 | 1.6 |
| Cole | 2023-Jan-01 | 6:46:42 | 1.0 |
| Cole | 2023-Jan-01 | 6:47:27 | 1.5 |
| Cole | 2023-Jan-01 | 6:47:49 | 2.2 |
| Cole | 2023-Jan-01 | 6:48:34 | 2.9 |
| Cole | 2023-Jan-01 | 6:49:04 | 3.3 |
| McNaught | 2023-Jan-01 | 12:28:03 | 1.8 |
| McNaught | 2023-Jan-01 | 12:28:38 | 2.4 |
| Cole | 2023-Jan-02 | 6:27:17 | 1.6 |
| Cole | 2023-Jan-02 | 6:27:36 | 1.1 |
| Cole | 2023-Jan-02 | 6:27:55 | 1.1 |
| Cole | 2023-Jan-02 | 6:28:23 | 1.0 |
| Cole | 2023-Jan-02 | 6:28:30 | 1.5 |
| Cole | 2023-Jan-02 | 6:28:50 | 1.5 |
| Cole | 2023-Jan-02 | 6:29:01 | 2.0 |
| Cole | 2023-Jan-02 | 6:29:20 | 2.2 |
| Cole | 2023-Jan-02 | 6:29:38 | 2.8 |
| Cole | 2023-Jan-02 | 6:30:57 | 5.3 |
| McNaught | 2023-Jan-03 | 11:51:00 | 2.8 |
| McNaught | 2023-Jan-03 | 11:52:45 | 2.7 |
| McNaught | 2023-Jan-03 | 11:51:39 | 3.1 |
| McNaught | 2023-Jan-04 | 11:32:15 | 3.6 |
| McNaught | 2023-Jan-04 | 11:32:54 | 3.5 |
| McNaught | 2023-Jan-04 | 11:34:00 | 2.8 |
| McNaught | 2023-Jan-04 | 11:34:52 | 2.2 |
| McNaught | 2023-Jan-06 | 10:54:30 | 4.9 |
| McNaught | 2023-Jan-06 | 10:55:17 | 4.3 |
| McNaught | 2023-Jan-06 | 10:55:40 | 4.4 |
| McNaught | 2023-Jan-06 | 10:56:15 | 3.8 |
| McNaught | 2023-Jan-06 | 10:57:02 | 3.0 |
| McNaught | 2023-Jan-06 | 10:57:16 | 2.5 |
| McNaught | 2023-Jan-06 | 10:58:23 | 2.0 |
| McNaught | 2023-Jan-06 | 10:58:52 | 2.1 |
| McNaught | 2023-Jan-07 | 10:35:44 | 6.0 |
| McNaught | 2023-Jan-07 | 10:35:57 | 5.7 |
| McNaught | 2023-Jan-07 | 10:36:20 | 5.4 |
| McNaught | 2023-Jan-07 | 10:36:49 | 5.5 |
| McNaught | 2023-Jan-07 | 10:37:09 | 4.8 |
| McNaught | 2023-Jan-07 | 10:37:37 | 3.6 |
| McNaught | 2023-Jan-07 | 10:38:27 | 3.0 |
| McNaught | 2023-Jan-07 | 10:38:35 | 3.2 |
| McNaught | 2023-Jan-07 | 10:38:45 | 2.6 |
| McNaught | 2023-Jan-07 | 10:40:07 | 1.6 |
| McNaught | 2023-Jan-07 | 10:40:24 | 1.4 |
| McNaught | 2023-Jan-07 | 10:40:40 | 1.8 |
| McNaught | 2023-Jan-07 | 10:41:06 | 2.5 |
| Harrington | 2023-Jan-09 | 12:16:30 | 4.4 |
| Harrington | 2023-Jan-09 | 12:21:00 | 5.5 |
| Harrington | 2023-Jan-10 | 12:00:30 | 4.5 |
| Harrington | 2023-Jan-10 | 12:01:30 | 4.7 |
| Harrington | 2023-Jan-10 | 12:03:00 | 5.5 |
| Harrington | 2023-Jan-10 | 12:04:00 | 6.7 |
| Harrington | 2023-Jan-14 | 12:23:00 | 3.0 |
| Harrington | 2023-Jan-14 | 12:24:00 | 2.1 |
| Harrington | 2023-Jan-14 | 12:25:00 | 2.0 |
| Harrington | 2023-Jan-14 | 12:26:00 | 3.1 |
| Harrington | 2023-Jan-14 | 12:26:30 | 3.3 |
| Harrington | 2023-Jan-14 | 12:28:00 | 5.5 |
| Mallama | 2023-Jan-15 | 10:29:55 | 0.8 |
| Mallama | 2023-Jan-15 | 10:30:12 | 1.0 |
| Mallama | 2023-Jan-15 | 10:30:20 | 1.7 |
| Mallama | 2023-Jan-15 | 10:30:30 | 2.0 |
| Mallama | 2023-Jan-15 | 10:30:40 | 2.4 |
| Mallama | 2023-Jan-15 | 10:30:55 | 2.7 |
| Mallama | 2023-Jan-15 | 10:31:06 | 3.0 |
| Mallama | 2023-Jan-15 | 10:31:20 | 3.5 |
| Mallama | 2023-Jan-15 | 10:31:41 | 4.0 |
| Harrington | 2023-Jan-15 | 12:05:30 | 2.0 |
| Harrington | 2023-Jan-15 | 12:06:00 | 1.9 |
| Harrington | 2023-Jan-15 | 12:07:30 | 3.0 |
| Harrington | 2023-Jan-15 | 12:08:00 | 3.3 |
| Harrington | 2023-Jan-15 | 12:09:30 | 4.7 |
| Mallama | 2023-Jan-24 | 23:04:20 | 3.5 |
| Mallama | 2023-Jan-24 | 23:05:10 | 2.1 |
| Mallama | 2023-Jan-24 | 23:05:45 | 3.0 |
| Mallama | 2023-Jan-25 | 0:41:50 | 2.9 |
| Mallama | 2023-Jan-25 | 0:42:10 | 2.6 |
| Respler | 2023-Jan-27 | 0:06:22 | 3.1 |
| Mallama | 2023-Jan-27 | 23:46:10 | 2.9 |
| Mallama | 2023-Jan-27 | 23:46:55 | 3.5 |
| Mallama | 2023-Jan-27 | 23:48:00 | 3.9 |
| Harrington | 2023-Feb-10 | 0:35:45 | 4.8 |
| Harrington | 2023-Feb-10 | 0:37:00 | 4.5 |
| Harrington | 2023-Feb-10 | 0:38:30 | 3.2 |
| Mallama | 2023-Feb-11 | 7:19:08 | 4.2 |
| Mallama | 2023-Feb-11 | 7:19:20 | 3.8 |
| Mallama | 2023-Feb-11 | 7:19:38 | 2.5 |
| Mallama | 2023-Feb-11 | 7:20:07 | 2.2 |
| Mallama | 2023-Feb-11 | 7:20:23 | 2.1 |
| Mallama | 2023-Feb-11 | 7:20:38 | 2.0 |
| Mallama | 2023-Feb-11 | 7:20:49 | 1.8 |
| Mallama | 2023-Feb-11 | 7:20:57 | 1.3 |
| Harrington | 2023-Feb-12 | 1:36:30 | 5.4 |
| Harrington | 2023-Feb-12 | 1:37:45 | 2.6 |
| Mallama | 2023-Feb-13 | 23:24:02 | 6.2 |
| Harrington | 2023-Feb-14 | 1:00:00 | 7.5 |
| Harrington | 2023-Feb-14 | 1:00:45 | 7.0 |
| Harrington | 2023-Feb-14 | 1:01:30 | 5.4 |
| Harrington | 2023-Feb-15 | 0:40:25 | 6.3 |
| Harrington | 2023-Feb-28 | 11:51:00 | 5.5 |
| Harrington | 2023-Feb-28 | 11:53:00 | 4.9 |
| Mallama | 2023-Mar-01 | 9:59:00 | 4.9 |
| Mallama | 2023-Mar-01 | 9:59:40 | 4.1 |
| Mallama | 2023-Mar-01 | 10:01:15 | 6.1 |
| Mallama | 2023-Mar-05 | 10:21:10 | 2.8 |
| Cole | 2023-Mar-06 | 5:19:42 | 1.3 |
| Cole | 2023-Mar-06 | 5:20:25 | 2.1 |
| Mallama | 2023-Mar-06 | 10:02:25 | 2.6 |
| Mallama | 2023-Mar-20 | 10:29:00 | 5.4 |
| Harrington | 2023-Mar-27 | 1:28:30 | 7.3 |
| Harrington | 2023-Mar-27 | 1:29:20 | 6.0 |
| Harrington | 2023-Mar-27 | 1:30:40 | 5.4 |
| Harrington | 2023-Mar-27 | 1:32:10 | 5.0 |
| Mallama | 2023-Mar-27 | 1:30:45 | 7.3 |
| Mallama | 2023-Mar-27 | 1:32:25 | 4.7 |
| Creed | 2023-Mar-27 | 1:32:05 | 4.2 |
| Mallama | 2023-Mar-30 | 0:33:00 | 6.9 |
| Mallama | 2023-Mar-31 | 0:13:20 | 6.2 |
| Mallama | 2023-Apr-10 | 1:54:45 | 4.0 |
| Mallama | 2023-Apr-11 | 1:34:50 | 3.1 |
| Mallama | 2023-Apr-12 | 1:14:55 | 3.7 |
| Mallama | 2023-Apr-13 | 0:55:00 | 4.0 |
| Harrington | 2023-Apr-13 | 2:30:00 | 4.3 |
| Harrington | 2023-Apr-13 | 2:32:40 | 3.0 |
| Mallama | 2023-Apr-14 | 0:35:05 | 3.9 |
| Harrington | 2023-Apr-15 | 1:50:50 | 4.9 |
| Harrington | 2023-Apr-15 | 1:52:00 | 3.5 |
| Harrington | 2023-Apr-15 | 1:52:50 | 3.0 |
| Harrington | 2023-Apr-15 | 1:53:40 | 2.0 |
| Harrington | 2023-Apr-15 | 1:54:30 | 2.7 |
| Mallama | 2023-Apr-19 | 0:33:50 | 5.1 |
| Mallama | 2023-Apr-19 | 0:35:35 | 3.5 |